\documentclass[a4paper]{PoS}

\usepackage{textcomp}
\usepackage{amsfonts} 
\usepackage{amsmath} 
\usepackage{slashed}
\usepackage{pifont}
\usepackage{multirow,lscape}
\usepackage{array}
\usepackage{subcaption}
\usepackage{verbatim}
\usepackage{bm}
\usepackage{comment}
\usepackage{xcolor}
\usepackage{url} 

\graphicspath{{./fig/}}

\providecommand{\wbar}[1]{\overline#1}

\providecommand{\mate}[3]{\langle#1\lvert#2\rvert#3\rangle}

\renewcommand{\Re}{\mathrm{Re}\,}
\renewcommand{\Im}{\mathrm{Im}\,}

\definecolor{HLBlue}{HTML}{6599FF}
\definecolor{HLOrange}{HTML}{FF6600}

\newcommand{\BK}{\hat{B}_{K}}
\newcommand{\Vcb}{|V_{cb}|}
\newcommand{\Vub}{|V_{ub}|}
\newcommand{\Vus}{|V_{us}|}

\newcommand{\eps}{\varepsilon}

\newcommand{\epsK}{\varepsilon_{K}}

\newcommand{\BtoDstp}{\bar{B} \to D^{(\ast)} \ell \bar{\nu}}
\newcommand{\BtoDst}{\bar{B} \to D^\ast \ell \bar{\nu}}

\newcommand{\wlee}[1]{\textcolor{red}{#1}} 

\newcommand{\red}[1]{\textcolor{red}{#1}}

\title{2019 Update on $\epsK$ with lattice QCD inputs}

\ShortTitle{$\epsK$ with lattice QCD inputs}

\author{Jeehun Kim, Sunkyu Lee, \speaker{Weonjong Lee} \\
        Lattice Gauge Theory Research Center, CTP, and FPRD, \\
        Department of Physics and Astronomy, \\
        Seoul National University,
        Seoul 08826, South Korea\\
        E-mail: \email{wlee@snu.ac.kr}}

\author{Yong-Chull Jang\\
        Physics Department,
        Brookhaven National Laboratory,
        Upton, NY11973, USA}

\author{Jaehoon Leem\\
        School of Physics,
        Korea Institute for Advanced Study (KIAS),
        Seoul 02455, South Korea}

\author{Sungwoo Park\\
        Los Alamos National Laboratory,
        Theoretical Division T-2,
        Los Alamos, NM87545, USA}

\author{LANL-SWME Collaboration}

\abstract{ We present updated results for $\epsK$ determined directly
  from the standard model (SM) with lattice QCD inputs such as $\BK$,
  $\Vcb$, $\Vus$, $\xi_0$, $\xi_2$, $\xi_\text{LD}$, $f_K$, and $m_c$.
  We find that the standard model with exclusive $\Vcb$ and other
  lattice QCD inputs describes only 65\% of the experimental value of
  $|\epsK|$ and does not explain its remaining 35\%, which leads to a
  strong tension in $|\epsK|$ at the $4.6\sigma \sim 4.2\sigma$ level
  between the SM theory and experiment.  We also find that this
  tension disappears when we use the inclusive value of $\Vcb$
  obtained using the heavy quark expansion based on QCD sum rules.  }

\FullConference{37th International Symposium on
                Lattice Field Theory - Lattice2019\\
                16-22 June 2019\\
                Wuhan, China}

\begin{document}

\section{Introduction}
This paper is an update of our previous papers \cite{ Bailey:2018feb,
  Bailey:2015tba, Bailey:2018aks, Jang:2017ieg, Bailey:2015frw}.
Here, we present recent progress in determination of $|\epsK|$ with
updated inputs from lattice QCD.

\section{Input parameters: $\Vcb$}
\label{sec:Vcb}
In Table \ref{tab:Vcb} (\subref{tab:ex-Vcb}) and (\subref{tab:in-Vcb}), we present
updated results for exclusive $\Vcb$ and inclusive $\Vcb$
respectively.
In Table \ref{tab:Vcb} (\subref{tab:ex-Vcb}), we present results for exclusive
$\Vcb$ of BELLE \cite{Abdesselam:2018nnh} and BABAR
\cite{Dey:2019bgc}.
They reported results obtained using both CLN and BGL methods, which
turn out to be consistent with each other.

In Table.~\ref{tab:Vcb} (\subref{fig:CLN-BGL}), we plot time evolution
of the $\Vcb$ results for the CLN analysis (blue line) as well as the
BGL analysis (red line) for the $\BtoDst$ decays.
Here, the black cross symbols with label \textsf{Gambino} represent
results from Refs.~\cite{ Bigi:2017njr, Bigi:2017jbd,
  Gambino:2019sif}, respectively.
The brown square symbol with label \textsf{Grinstein} represents
results from Ref.~\cite{Grinstein:2017nlq}.
The green circle symbol with label \textsf{BELLE-17} represents
results from Ref.~\cite{Abdesselam:2017kjf}.
The magenta circle symbols with label \textsf{BELLE-18} represent
results from Ref.~\cite{Abdesselam:2018nnh}.
The green triangle symbols with label \textsf{BELLE-19} represent
results from Ref.~\cite{BELLE:2019}.
The orange diamond symbols with label \textsf{BABAR} represents
results from Ref.~\cite{Dey:2019bgc}.
In 2017 when Bigi, Gambino, and Schacht \cite{ Bigi:2017njr,
  Grinstein:2017nlq} first raised a claim that there might be an
inconsistency in exclusive $\Vcb$ between the CLN and BGL analyses on
the BELLE-2017 tagged data set of the $\BtoDst$ decays, the BGL
results seemed to be superficially consistent with those for inclusive
$\Vcb$.
However, the 2019 analyses of both BELLE \cite{ Abdesselam:2018nnh}
(on the untagged data) and BABAR \cite{ Dey:2019bgc} show that the
results of the BGL analysis might be consistent with those of the CLN
analysis, which denies the previous claim of Refs.~\cite{
  Bigi:2017njr, Grinstein:2017nlq}.
The pink dashed line with label \textsf{FLAG-19} represents
preliminary results of BELLE-18 which the FLAG 2019 report took over
to do their analysis on $\Vcb$.
Hence, if you are to use the $\Vcb$ results of FLAG 2019, please do it
with proper caution, since they might be out of date.
The green dashed line with label \textsf{SWME-19} represents results
of BELLE-19 \cite{ Abdesselam:2018nnh} which we use for the analysis
on $\epsK$ in this paper.

In Table \ref{tab:Vcb} (\subref{fig:Vcb-Vub}), we show the plot made
by HFLAV.
Results on this plot are available on the web \cite{ HFLAV:2019}, but
not published in any journal yet.
Recently, there has been an interesting claim that the $\Vcb$ puzzle
might be resolved if we include $\mathcal{O}(1/m_c^2)$ corrections
in the data analysis \cite{Bordone:2019vic}.
%
%
%
%
%

\begin{table}[t!]
  \begin{subtable}{0.55\linewidth}
    \renewcommand{\arraystretch}{1.25}
    \center
    \vspace*{-7mm}
    \resizebox{0.99\textwidth}{!}{
      \begin{tabular}{l l l l}
        \hline\hline
        channel & method & value & Ref. \\ \hline
        combined  &     & $39.13(59)$          & HFLAV-17 \cite{Amhis:2016xyh}
        \\
        combined  &     & \red{$39.25(56)$}    & HFLAV-19 \cite{HFLAV:2019}
        \\ \hline
        $\BtoDst$ & CLN & \red{38.4(2)(6)(6)}  & BELLE-19 \cite{Abdesselam:2018nnh}
        \\
        $\BtoDst$ & BGL & \red{38.3(3)(7)(6)}  & BELLE-19 \cite{Abdesselam:2018nnh}
        \\ \hline
        $\BtoDst$ & CLN & \red{38.40(84)}      & BABAR-19 \cite{Dey:2019bgc}
        \\
        $\BtoDst$ & BGL & \red{38.36(90)}      & BABAR-19 \cite{Dey:2019bgc}
        \\ \hline\hline
      \end{tabular}
    } 
    \caption{Exclusive $|V_{cb}|$ in units of $10^{-3}$.}
    \label{tab:ex-Vcb}
  \end{subtable}
  \hfill
  \begin{subfigure}{0.45\linewidth}
    \vspace*{-7mm}
    \centering
    \includegraphics[width=1.1\textwidth]{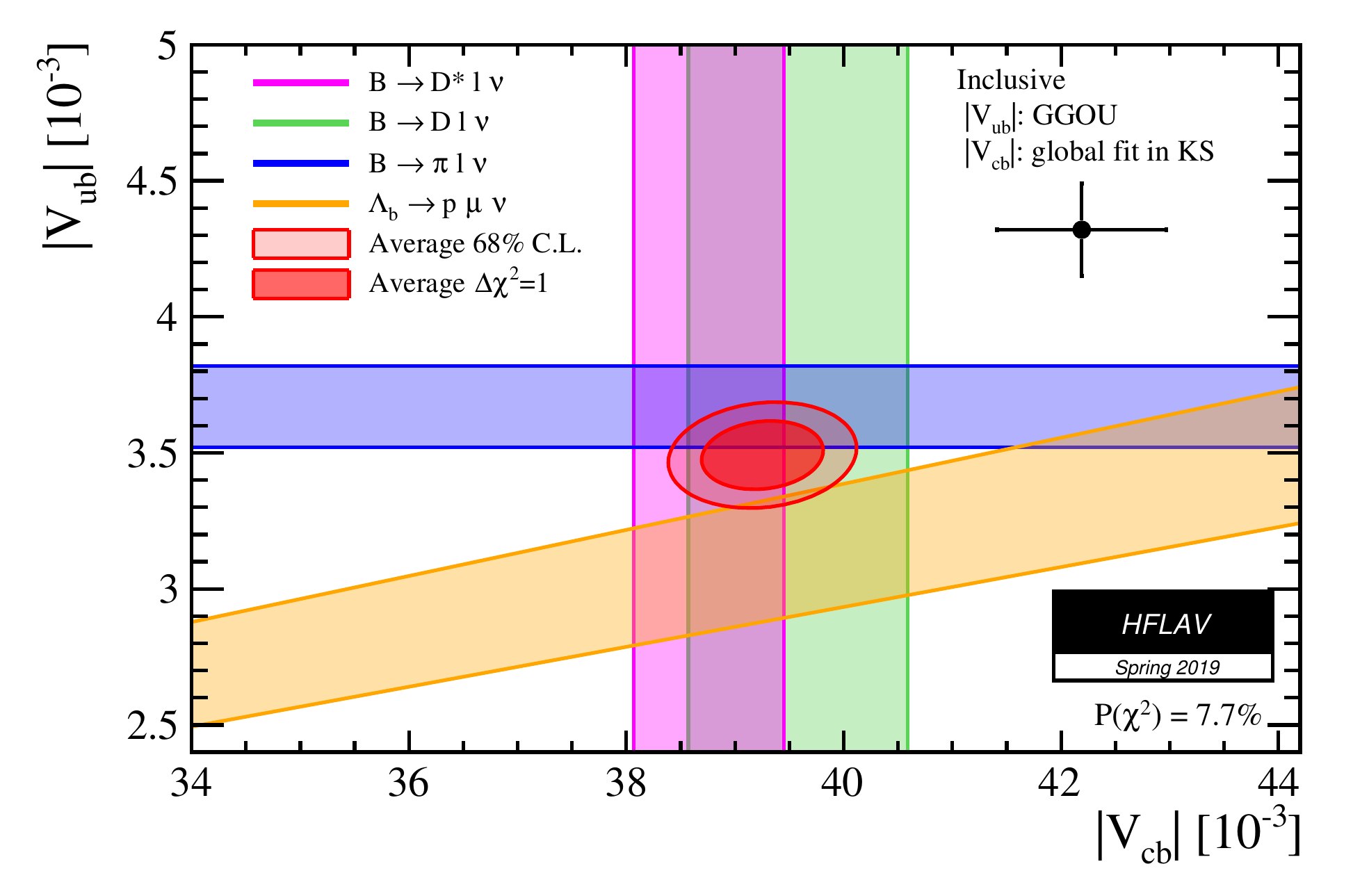}
    \caption{$|V_{cb}|$ versus $|V_{ub}|$.}
    \label{fig:Vcb-Vub}
  \end{subfigure}
  \begin{subfigure}{0.67\linewidth}
    \centering
    \includegraphics[width=1.0\textwidth]{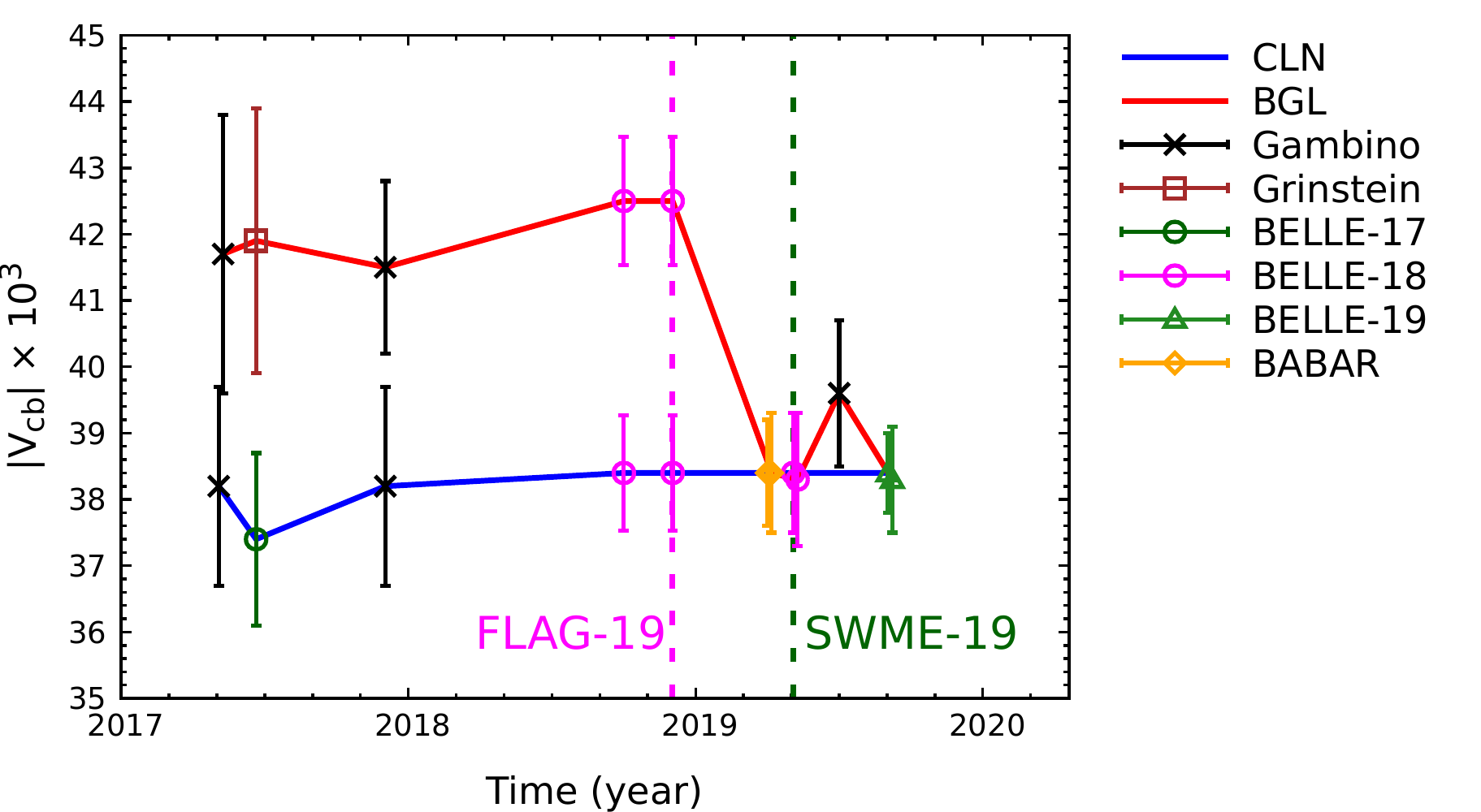}
    \caption{CLN versus BGL.}
    \label{fig:CLN-BGL}
  \end{subfigure}
  \hfill
  \begin{subtable}{0.32\linewidth}
    \renewcommand{\arraystretch}{1.4}
    \center
    \resizebox{0.99\textwidth}{!}{
      \begin{tabular}{l l c}
        \hline\hline
        channel & value & Ref. \\ \hline
        kinetic scheme & $42.19(78)$ & \cite{Amhis:2016xyh}
        \\
        1S scheme      & $41.98(45)$ & \cite{Amhis:2016xyh}
        \\ \hline\hline
      \end{tabular}
    } 
    \caption{Inclusive $|V_{cb}|$ in units of $10^{-3}$.}
    \label{tab:in-Vcb}
    \vspace*{5mm}
    \resizebox{0.99\linewidth}{!}{
      \begin{tabular}{lcc}
        \hline\hline
        method & $\xi_0$ & Ref. \\ \hline
        indirect & $-1.63(19) \times 10^{-4}$ & \cite{Blum:2015ywa}
        \\
        direct  & $-0.57(49) \times 10^{-4}$  & \cite{Bai:2015nea}
        \\ \hline\hline
      \end{tabular}
    } 
    \caption{Results for $\xi_0$.}
    \label{tab:xi0}
  \end{subtable}
  
  \caption{ Results for $\Vcb$: (\subref{tab:ex-Vcb}) exclusive
    $\Vcb$, (\subref{fig:Vcb-Vub}) $\Vcb$ versus $\Vub$,
    (\subref{fig:CLN-BGL}) Time evolution for exclusive $\Vcb$
    obtained using CLN and BGL, and (\subref{tab:in-Vcb}) inclusive
    $\Vcb$; (\subref{tab:xi0}) results for $\xi_0$.}
  \label{tab:Vcb}
\end{table}

\section{Input parameter $\xi_0$}
The absorptive part of long distance effects on $\epsK$ is parametrized
into $\xi_0$.
\begin{align}
  \xi_0  &= \frac{\Im A_0}{\Re A_0}, \qquad
  \xi_2 = \frac{\Im A_2}{\Re A_2}, \qquad
  \Re \left(\frac{\eps'}{\eps} \right) =
  \frac{\omega}{\sqrt{2} |\eps_K|} (\xi_2 - \xi_0) \,.
  \label{eq:e'/e:xi0}
\end{align}
There are two independent methods to determine $\xi_0$ in lattice QCD:
the indirect and direct methods.
The indirect method is to determine $\xi_0$ using
Eq.~\eqref{eq:e'/e:xi0} with lattice QCD results for $\xi_2$ combined
with experimental results for $\eps'/\eps$, $\epsK$, and $\omega$.
The direct method is to determine $\xi_0$ directly using the lattice
QCD results for $\Im A_0$, combined with experimental results for $\Re
A_0$.
In Table~\ref{tab:Vcb} (\subref{tab:xi0}), we summarize results for
$\xi_0$ calculated by RBC-UKQCD.
Here, we use the results of the indirect method for $\xi_0$ to
evaluate $\epsK$, since its systematic and statistical errors are much
smaller.

\section{Input parameters: Wolfenstein parameters, $\BK$,
  $\xi_\text{LD}$, and others}

\begin{table}[h!]
%
  \begin{subtable}{0.6\linewidth}
    \renewcommand{\arraystretch}{1.2}
    \vspace*{-4mm}
    \resizebox{0.99\linewidth}{!}{
      \begin{tabular}{ c | l c | l c | l c }
        \hline\hline
        WP
        & \multicolumn{2}{c|}{CKMfitter}
        & \multicolumn{2}{c|}{UTfit}
        & \multicolumn{2}{c}{AOF}
        \\ \hline
        $\lambda$
        & $0.22475(25)$        & \cite{Charles:2004jd}
        & $0.22500(100)$       & \cite{Bona:2006ah}
        & \wlee{ $0.2243(5)$ } & \cite{Tanabashi:2018oca}
        \\ \hline
        $\bar{\rho}$
        & $0.1577(96)$ & \cite{Charles:2004jd}
        & $0.148(13)$  & \cite{Bona:2006ah}
        & $0.146(22)$  & \cite{Martinelli:2017}
        \\ \hline
        $\bar{\eta}$
        & $0.3493(95)$ & \cite{Charles:2004jd}
        & $0.348(10)$  & \cite{Bona:2006ah}
        & $0.333(16)$  & \cite{Martinelli:2017}
        \\ \hline\hline
      \end{tabular}
    } 
    \caption{Wolfenstein parameters}
    \label{tab:WP}
  \end{subtable} 
  \hfill
  \begin{subtable}{0.23\linewidth}
    \renewcommand{\arraystretch}{1.2}
    \vspace*{-4mm}
    \resizebox{0.99\linewidth}{!}{
      \begin{tabular}[b]{ c l c }
        \hline\hline
        Input & Value & Ref.
        \\ \hline
        $\eta_{cc}$ & $1.72(27)$   & \cite{Bailey:2015tba}
        \\
        $\eta_{tt}$ & $0.5765(65)$ & \cite{Buras2008:PhysRevD.78.033005}
        \\
        $\eta_{ct}$ & $0.496(47)$  & \cite{Brod2010:prd.82.094026}
        \\ \hline\hline
      \end{tabular}
    } 
    \caption{$\eta_{ij}$}
    \label{tab:eta}
  \end{subtable} 
  \caption{ (\subref{tab:WP}) Wolfenstein parameters and
    (\subref{tab:eta}) QCD corrections: $\eta_{ij}$ with $i,j = c,t$.}
  \label{tab:input-WP-eta}
\end{table}

In Table \ref{tab:input-WP-eta}\,(\subref{tab:WP}), we present the
Wolfenstein parameters on the market.
As explained in Ref.~\cite{ Bailey:2018feb, Bailey:2015frw}, we use
the results of angle-only-fit (AOF) in Table
\ref{tab:input-WP-eta}\,(\subref{tab:WP}) in order to avoid unwanted
correlation between $(\epsK, \Vcb)$, and $(\bar\rho, \bar\eta)$.
We determine $\lambda$ from $\Vus$ which is obtained from the $K_{\ell
  2}$ and $K_{\ell 3}$ decays using lattice QCD inputs for form
factors and decay constants.
We determine the $A$ parameter from $\Vcb$.

In FLAG 2019 \cite{ Aoki:2019cca}, they report lattice QCD results for
$\BK$ with $N_f=2$, $N_f=2+1$, and $N_f = 2+1+1$.
Here, we use the results for $\BK$ with $N_f=2+1$, which is obtained
by taking an average over the four data points from BMW 11, Laiho 11
RBC-UKQCD 14, and SWME 15 in Table
\ref{tab:input-BK-other}\;(\subref{tab:BK}).

\begin{table}[t!]
%
  \begin{subtable}{0.40\linewidth}
    \renewcommand{\arraystretch}{1.45}
    \vspace*{-5mm}
    \resizebox{1.0\linewidth}{!}{
      \begin{tabular}{ l  c  l }
        \hline\hline
        Collaboration & Ref. & $\BK$  \\ \hline
        SWME 15       & \cite{Jang:2015sla} & $0.735(5)(36)$     \\
        RBC/UKQCD 14  & \cite{Blum:2014tka} & $0.7499(24)(150)$  \\
        Laiho 11      & \cite{Laiho:2011np} & $0.7628(38)(205)$  \\
        BMW 11        & \cite{Durr:2011ap}  & $0.7727(81)(84)$  \\ \hline
        FLAG 2019     & \cite{Aoki:2019cca} & $0.7625(97)$
        \\ \hline\hline
      \end{tabular}
    } 
    \caption{$\BK$}
    \label{tab:BK}
  \end{subtable} 
  \hfill
  \begin{subtable}{0.50\linewidth}
    \renewcommand{\arraystretch}{1.2}
    \vspace*{-5mm}
    \resizebox{1.0\linewidth}{!}{
      \begin{tabular}{ c l c }
        \hline\hline
        Input & Value & Ref. \\ \hline
        $G_{F}$
        & $1.1663787(6) \times 10^{-5}$ GeV$^{-2}$
        & PDG-19 \cite{Tanabashi:2018oca} \\ \hline
        $M_{W}$
        & \wlee{$80.379(12)$} GeV
        & PDG-19 \cite{Tanabashi:2018oca}\\ \hline
        $\theta$
        & $43.52(5)^{\circ}$
        & PDG-19 \cite{Tanabashi:2018oca} \\ \hline
        $m_{K^{0}}$
        & $497.611(13)$ MeV
        & PDG-19 \cite{Tanabashi:2018oca} \\ \hline
        $\Delta M_{K}$
        & $3.484(6) \times 10^{-12}$ MeV
        & PDG-19 \cite{Tanabashi:2018oca} \\ \hline
        $F_K$
        & \wlee{ $155.7(3)$ } MeV
        & FLAG-19 \cite{Aoki:2019cca}
        \\ \hline\hline
      \end{tabular}
    } 
    \caption{Other parameters}
    \label{tab:other}
  \end{subtable} 
  \caption{ (\subref{tab:BK}) Results for $\BK$ and
    (\subref{tab:other}) other input parameters.}
  \label{tab:input-BK-other}
\end{table}

The dispersive long distance (LD) effect is defined as
\begin{align}
  \xi_\text{LD} &=  \frac{m^\prime_\text{LD}}{\sqrt{2} \Delta M_K} \,,
  \qquad
  m^\prime_\text{LD}
  = -\Im \left[ \mathcal{P}\sum_{C}
    \frac{\mate{\wbar{K}^0}{H_\text{w}}{C} \mate{C}{H_\text{w}}{K^0}}
         {m_{K^0}-E_{C}}  \right]
  \label{eq:xi-LD}
\end{align}
As explained in Refs.~\cite{ Bailey:2018feb}, there are two
independent methods to estimate $\xi_\text{LD}$: one is the BGI
estimate \cite{ Buras:2010}, and the other is the RBC-UKQCD estimate
\cite{ Christ:2012, Christ:2014qwa}.
The BGI method is to estimate the size of $\xi_\text{LD}$ using
chiral perturbation theory as follows,
\begin{align}
  \xi_\text{LD} &= -0.4(3) \times \frac{\xi_0}{ \sqrt{2} }
  \label{eq:xiLD:bgi}
\end{align}
The RBC-UKQCD method is to estimate the size of $\xi_\text{LD}$
as follows,
\begin{align}
  \xi_\text{LD} &= (0 \pm 1.6)\%.
  \label{eq:xiLD:rbc}
\end{align}
Here, we use both methods to estimate the size of $\xi_\text{LD}$.

In Table \ref{tab:input-WP-eta}\;(\subref{tab:eta}), we present higher
order QCD corrections: $\eta_{ij}$ with $i,j = t,c$.
A new approach using $u-t$ unitarity instead of $c-t$ unitarity
appeared in Ref.~\cite{ Brod:2019rzc}, which is supposed to have a
better convergence with respect to the charm quark mass.
Here, we have not incorporated this into our analysis yet, but will
do it in near future.

In Table \ref{tab:input-BK-other}\;(\subref{tab:other}), we present other
input parameters needed to evaluate $\epsK$.
Here, the $W$ boson mass $M_W$ and the kaon decay constant $F_K$ have
been updated since Lattice 2018.
In Table \ref{tab:m_c:m_t}, we present the charm quark mass $m_c(m_c)$
and top quark mass $m_t(m_t)$.
From FLAG 2019 \cite{ Aoki:2019cca}, we take the results for $m_c
(m_c)$ with $N_f = 2+1$, since there is some discrepancy in those with
$N_f = 2+1+1$.
For the top quark mass, we use the PDG 2019 results to obtain $m_t
(m_t)$.

\begin{table}[t!]
  \begin{subtable}{0.40\linewidth}
    \vspace*{-5mm}
    \renewcommand{\arraystretch}{1.2}
    \resizebox{0.99\linewidth}{!}{
      \begin{tabular}{ l l l l }
        \hline\hline
        {\footnotesize Collaboration} & $N_f$ & $m_c(m_c)$ & Ref.
        \\ \hline
        FLAG 2019       & $2+1$   & \wlee{$1.275(5)$}  & \cite{Aoki:2019cca}
        \\
        FLAG 2019       & $2+1+1$ & $1.280(13)$        & \cite{Aoki:2019cca}
        \\ \hline\hline
      \end{tabular}
    } 
    \caption{$m_c(m_c)$ [GeV]}
    \label{tab:m_c}
  \end{subtable} 
  \hfill
  \begin{subtable}{0.50\linewidth}
    \renewcommand{\arraystretch}{1.2}
    \vspace*{-5mm}
    \resizebox{0.99\linewidth}{!}{
      \begin{tabular}{ l l l l }
        \hline\hline
        {\footnotesize Collaboration} & $M_t$ & $m_t(m_t)$ & Ref.
        \\ \hline
        PDG 2018      & $173.0 \pm 0.4$
        & $163.17 \pm 0.38 \pm 0.17$
        & \cite{Tanabashi:2018oca} 
        \\
        PDG 2019      & $172.9 \pm 0.4$
        & \wlee{ $163.08 \pm 0.38 \pm 0.17$ }
        & \cite{Tanabashi:2018oca}
       \\ \hline\hline
      \end{tabular}
    } 
    \caption{$m_t(m_t)$ [GeV]}
    \label{tab:m_t}
  \end{subtable} 
  \caption{  Results for (\subref{tab:m_c}) charm quark mass and
    (\subref{tab:m_t}) top quark mass. }
  \label{tab:m_c:m_t}
\end{table}

\section{Results for $\epsK$}

\begin{figure}[t!]
  \begin{subfigure}{0.47\linewidth}
    \vspace*{-3mm}
    \includegraphics[width=1.0\linewidth]
       {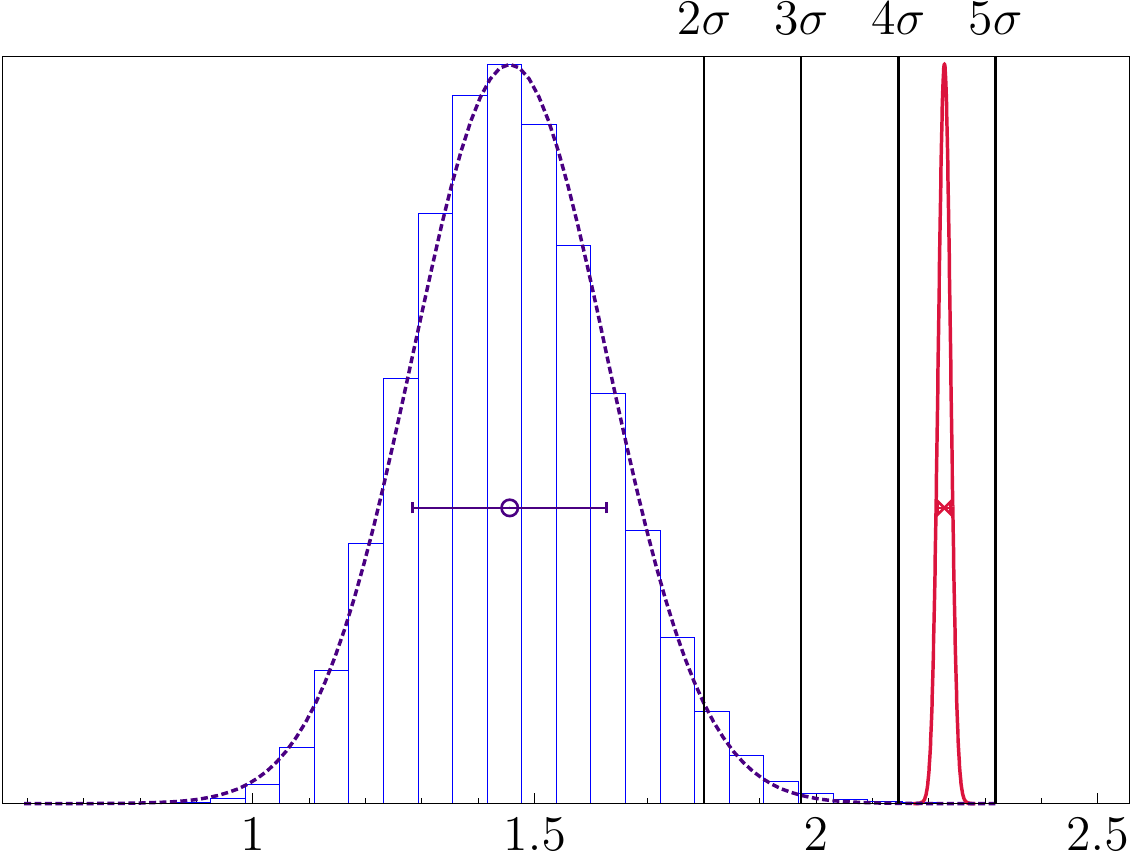}
    \caption{Exclusive $\Vcb$ (BELLE-19, CLN)}
    \label{fig:epsK-ex:rbc}
  \end{subfigure}
  \hfill
  \begin{subfigure}{0.47\linewidth}
    \vspace*{-3mm}
    \includegraphics[width=1.0\linewidth]
                    {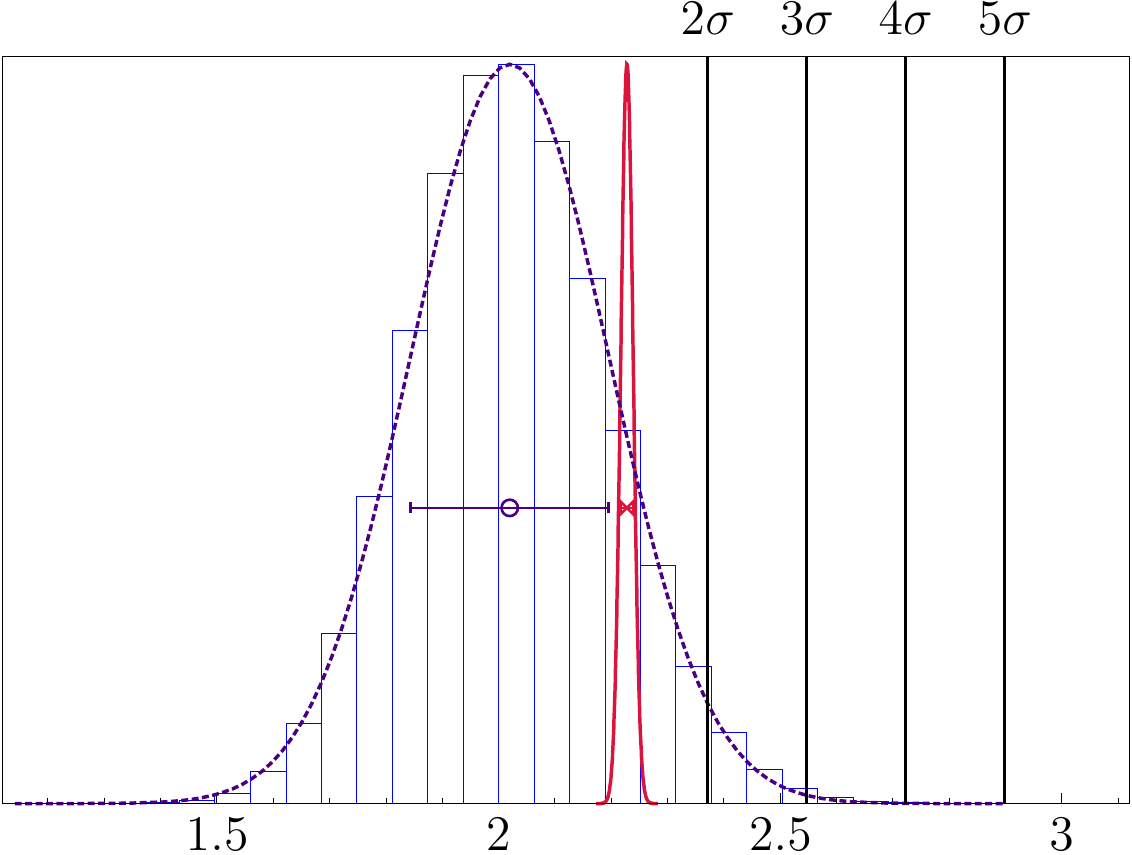}
    \caption{Inclusive $\Vcb$ (1S scheme)}
    \label{fig:epsK-in:rbc}
  \end{subfigure}
  \caption{$|\epsK|$ with (\subref{fig:epsK-ex:rbc}) exclusive $\Vcb$
    (left) and (\subref{fig:epsK-in:rbc}) inclusive $\Vcb$ (right) in
    units of $1.0\times 10^{-3}$.}
  \label{fig:epsK:cmp:rbc}
\end{figure}

In Fig.~\ref{fig:epsK:cmp:rbc}, we show results for $|\epsK|$
evaluated directly from the standard model (SM) with lattice QCD
inputs given in the previous sections.
In Fig.~\ref{fig:epsK:cmp:rbc}\;(\subref{fig:epsK-ex:rbc}), the blue
curve represents the theoretical evaluation of $|\epsK|$ obtained
using the FLAG-2019 results for $\BK$, AOF for Wolfenstein parameters,
the (BELLE-19, CLN) results for exclusive $\Vcb$, and the RBC-UKQCD
estimate for $\xi_\text{LD}$.
The red curve in Fig.~\ref{fig:epsK:cmp:rbc} represents the experimental
results for $|\epsK|$.
In Fig.~\ref{fig:epsK:cmp:rbc}\;(\subref{fig:epsK-in:rbc}), the blue
curve represents the same as in
Fig.~\ref{fig:epsK:cmp:rbc}\;(\subref{fig:epsK-ex:rbc}) except for
using the 1S scheme results for the inclusive $\Vcb$.

Our results for $|\epsK|^\text{SM}$ are summarized in Table
\ref{tab:epsK}.
Here, the superscript ${}^\text{SM}$ represents the theoretical
expectation value of $|\epsK|$ obtained directly from the SM.
The superscript ${}^\text{Exp}$ represents the experimental value
of $|\epsK| = 2.228(11) \times 10^{-3}$.
Results in Table \ref{tab:epsK}\;(\subref{tab:epsK:rbc}) are obtained
using the RBC-UKQCD estimate for $\xi_\text{LD}$, and those in
Table \ref{tab:epsK}\;(\subref{tab:epsK:bgi}) are obtained using
the BGI estimate for $\xi_\text{LD}$.
In Table \ref{tab:epsK}\;(\subref{tab:epsK:rbc}), we find that the
theoretical expectation values of $|\epsK|^\text{SM}$ with lattice QCD
inputs (with exclusive $\Vcb$) has $4.6\sigma \sim 4.2\sigma$ tension
with the experimental value of $|\epsK|^\text{Exp}$, while there is no
tension with inclusive $\Vcb$ (obtained using heavy quark expansion and
QCD sum rules).

\begin{table}[tbhp]
%
  \begin{subtable}{1.0\linewidth}
    \center
    \renewcommand{\arraystretch}{1.2}
    \resizebox{0.7\linewidth}{!}{
      \begin{tabular}{@{\qquad} l @{\qquad} l @{\qquad} l @{\qquad} l @{\qquad} l @{\qquad} }
        \hline\hline
        $\Vcb$    & method   & reference & $|\epsK|^\text{SM}$ & $\Delta\epsK$
        \\ \hline
        exclusive & CLN      & BELLE-19 & $1.456 \pm 0.172$ & $4.47\sigma$
        \\
        exclusive & BGL      & BELLE-19 & $1.443 \pm 0.181$ & $4.32\sigma$
        \\ \hline
        exclusive & CLN      & BABAR-19 & $1.456 \pm 0.169$ & $4.55\sigma$
        \\
        exclusive & BGL      & BABAR-19 & $1.451 \pm 0.175$ & $4.44\sigma$
        \\ \hline
        exclusive & combined & HFLAV-19 & $1.576 \pm 0.154$ & $4.23\sigma$
        \\ \hline
        inclusive & kinetic  & HFLAV-17 & $2.060 \pm 0.212$ & $0.79\sigma$
        \\
        inclusive & 1S       & HFLAV-17 & $2.020 \pm 0.176$ & $1.18\sigma$
        \\ \hline\hline
      \end{tabular}
    } 
    \caption{RBC-UKQCD estimate for $\xi_\text{LD}$}
    \label{tab:epsK:rbc}
  \end{subtable} 
  \begin{subtable}{1.0\linewidth}
    \vspace*{3mm}
    \center
    \renewcommand{\arraystretch}{1.2}
    \resizebox{0.7\linewidth}{!}{
      \begin{tabular}{@{\qquad} l @{\qquad} l @{\qquad} l @{\qquad} l @{\qquad} l @{\qquad} }
        \hline\hline
        $\Vcb$    & method   & reference  & $|\epsK|^\text{SM}$ & $\Delta\epsK$
        \\ \hline
        exclusive & CLN      & BELLE-19 & $1.501 \pm 0.174$ & $4.16\sigma$
        \\
        exclusive & BGL      & BELLE-19 & $1.488 \pm 0.183$ & $4.04\sigma$
        \\ \hline\hline
      \end{tabular}
    } 
    \caption{BGI estimate for $\xi_\text{LD}$}
    \label{tab:epsK:bgi}
  \end{subtable} 
  \caption{ $|\epsK|$ in units of $1.0\times 10^{-3}$, and
    $\Delta\epsK = |\epsK|^\text{Exp} - |\epsK|^\text{SM}$.}
  \label{tab:epsK}
\end{table}

\begin{figure}[htbp]
  \begin{subfigure}{0.470\linewidth}
    \includegraphics[width=\linewidth]
                    {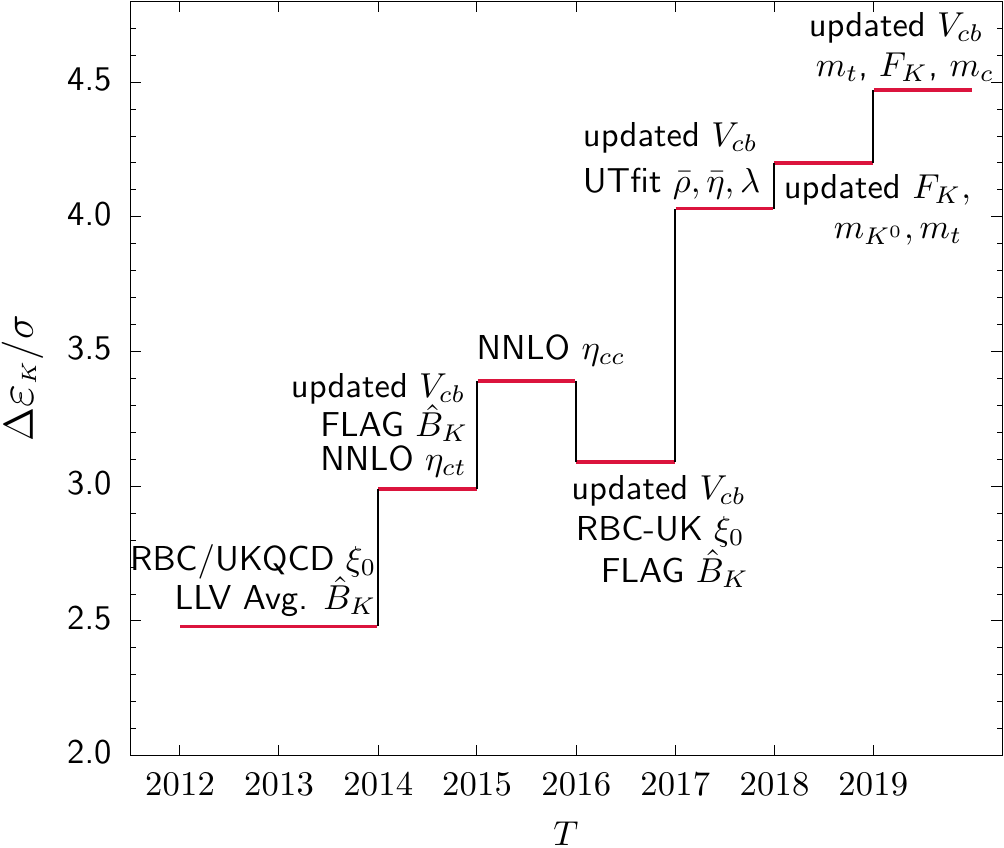}
    \caption{Time evolution of $\Delta \epsK/\sigma$}
    \label{fig:depsK:rbc:his}
  \end{subfigure}
  \hfill
  \begin{subfigure}{0.505\linewidth}
    \includegraphics[width=\linewidth]
                    {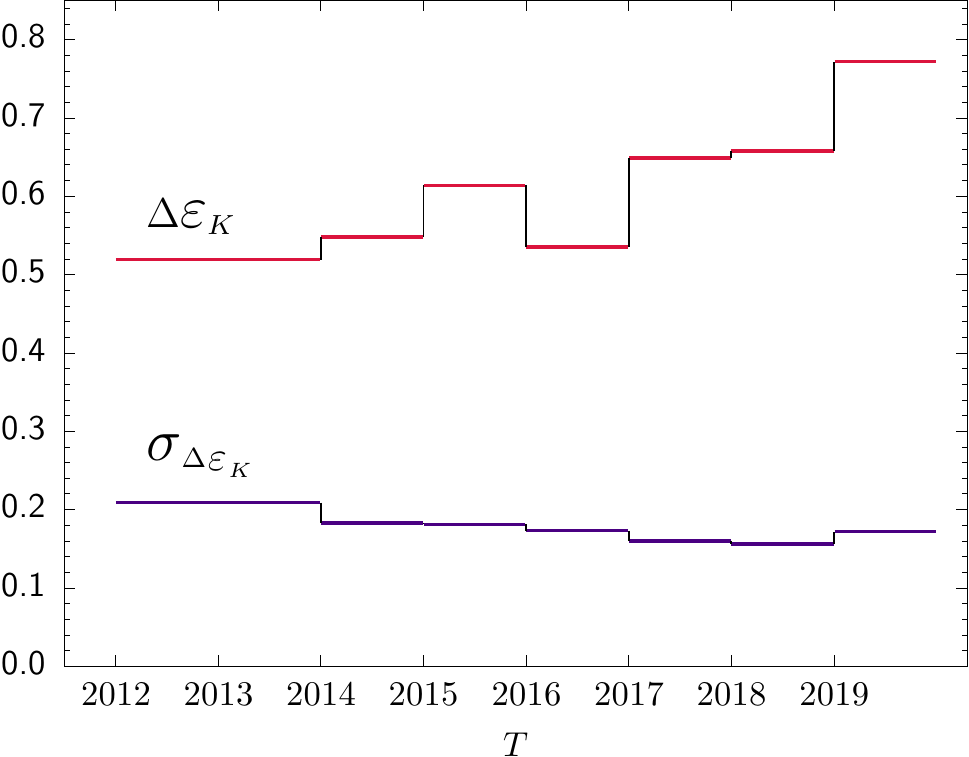}
    \caption{Time evolution of the average and error of $\Delta\epsK$}
    \label{fig:depsK+sigma:rbc:his}
  \end{subfigure}
  \caption{ Time history of (\subref{fig:depsK:rbc:his})
    $\Delta\epsK/\sigma$, and (\subref{fig:depsK+sigma:rbc:his})
    $\Delta\epsK$ and $\sigma_{\Delta\epsK}$. }
  \label{fig:depsK:sum:rbc:his}
\end{figure}

In Fig.~\ref{fig:depsK:sum:rbc:his}\;(\subref{fig:depsK:rbc:his}), we
show the time evolution of $\Delta\epsK$ starting from 2012 to 2019.
In 2012, $\Delta\epsK$ was $2.5\sigma$, but now it is $4.5\sigma$
with exclusive $\Vcb$ (BELLE-19, CLN).
In Fig.~\ref{fig:depsK:sum:rbc:his}\;(\subref{fig:depsK+sigma:rbc:his}),
we show the time evolution of the average $\Delta\epsK$ and the error
$\sigma_{\Delta\epsK}$ during the period of 2012--2019.

At present, we find that the largest error ($\approx 50\%$) in
$|\epsK|^\text{SM}$ comes from $\Vcb$.
Hence, it is of crucial importance to reduce the error in $\Vcb$
significantly.
To achieve this goal, there is an on-going project to extract
exclusive $\Vcb$ using the Oktay-Kronfeld (OK) action for the heavy
quarks to calculate the form factors for $\BtoDstp$ decays \cite{
  Ben:2019, Seungyeob:2019, Bhattacharya:2018ibo, Bailey:2017xjk,
  Bailey:2017zgt}.

\acknowledgments
We thank Rajan Gupta, Andrew Lytle, and Martin Jung for helpful
discussion.
The research of W.~Lee is supported by the Mid-Career Research
Program (Grant No.~NRF-2019R1A2C2085685) of the NRF grant funded by
the Korean government (MOE).
This work was supported by Seoul National University Research Grant in
2019.
W.~Lee would like to acknowledge the support from the KISTI
supercomputing center through the strategic support program for the
supercomputing application research (No.~KSC-2017-G2-0009).
Computations were carried out on the DAVID cluster at Seoul National
University.

\bibliography{refs}

\providecommand{\href}[2]{#2}\begingroup\raggedright\begin{thebibliography}{10}

\bibitem{Bailey:2018feb}
J.~A. Bailey {\em et~al.} {\em Phys. Rev.} {\bf D98} (2018), no.~9 094505,
  [\href{http://xxx.lanl.gov/abs/1808.09657}{{\tt 1808.09657}}].

\bibitem{Bailey:2015tba}
J.~A. Bailey, Y.-C. Jang, W.~Lee, and S.~Park {\em Phys. Rev.} {\bf D92}
  (2015), no.~3 034510, [\href{http://xxx.lanl.gov/abs/1503.05388}{{\tt
  1503.05388}}].

\bibitem{Bailey:2018aks}
J.~A. Bailey {\em et~al.} {\em PoS} {\bf LATTICE2018} (2018) 284,
  [\href{http://xxx.lanl.gov/abs/1810.09761}{{\tt 1810.09761}}].

\bibitem{Jang:2017ieg}
Y.-C. Jang, W.~Lee, S.~Lee, and J.~Leem {\em EPJ Web Conf.} {\bf 175} (2018)
  14015, [\href{http://xxx.lanl.gov/abs/1710.06614}{{\tt 1710.06614}}].

\bibitem{Bailey:2015frw}
J.~A. Bailey, Y.-C. Jang, W.~Lee, and S.~Park {\em PoS} {\bf LATTICE2015}
  (2015) 348, [\href{http://xxx.lanl.gov/abs/1511.00969}{{\tt 1511.00969}}].

\bibitem{Abdesselam:2018nnh}
{\bf Belle} Collaboration, E.~Waheed {\em et~al.} {\em Phys. Rev.} {\bf D100}
  (2019), no.~5 052007, [\href{http://xxx.lanl.gov/abs/1809.03290}{{\tt
  1809.03290}}].

\bibitem{Dey:2019bgc}
{\bf BaBar} Collaboration, J.~P. Lees {\em et~al.} {\em Phys. Rev. Lett.} {\bf
  123} (2019), no.~9 091801, [\href{http://xxx.lanl.gov/abs/1903.10002}{{\tt
  1903.10002}}].

\bibitem{Bigi:2017njr}
D.~Bigi, P.~Gambino, and S.~Schacht {\em Phys. Lett.} {\bf B769} (2017)
  441--445, [\href{http://xxx.lanl.gov/abs/1703.06124}{{\tt 1703.06124}}].

\bibitem{Bigi:2017jbd}
D.~Bigi, P.~Gambino, and S.~Schacht {\em JHEP} {\bf 11} (2017) 061,
  [\href{http://xxx.lanl.gov/abs/1707.09509}{{\tt 1707.09509}}].

\bibitem{Gambino:2019sif}
P.~Gambino, M.~Jung, and S.~Schacht {\em Phys. Lett.} {\bf B795} (2019)
  386--390, [\href{http://xxx.lanl.gov/abs/1905.08209}{{\tt 1905.08209}}].

\bibitem{Grinstein:2017nlq}
B.~Grinstein and A.~Kobach {\em Phys. Lett.} {\bf B771} (2017) 359--364,
  [\href{http://xxx.lanl.gov/abs/1703.08170}{{\tt 1703.08170}}].

\bibitem{Abdesselam:2017kjf}
A.~Abdesselam {\em et~al.} \href{http://xxx.lanl.gov/abs/1702.01521}{{\tt
  1702.01521}}.

\bibitem{BELLE:2019}
BELLE. Private communication with Phillip Urquiso, 2019.

\bibitem{HFLAV:2019}
\url{https://hflav-eos.web.cern.ch/hflav-eos/semi/spring19/main.shtml}.

\bibitem{Bordone:2019vic}
M.~Bordone, M.~Jung, and D.~van Dyk
  \href{http://xxx.lanl.gov/abs/1908.09398}{{\tt 1908.09398}}.

\bibitem{Amhis:2016xyh}
Y.~Amhis {\em et~al.} {\em Eur. Phys. J.} {\bf C77} (2017), no.~12 895,
  [\href{http://xxx.lanl.gov/abs/1612.07233}{{\tt 1612.07233}}].

\bibitem{Blum:2015ywa}
T.~Blum {\em et~al.} {\em Phys. Rev.} {\bf D91} (2015), no.~7 074502,
  [\href{http://xxx.lanl.gov/abs/1502.00263}{{\tt 1502.00263}}].

\bibitem{Bai:2015nea}
Z.~Bai {\em et~al.} {\em Phys. Rev. Lett.} {\bf 115} (2015), no.~21 212001,
  [\href{http://xxx.lanl.gov/abs/1505.07863}{{\tt 1505.07863}}].

\bibitem{Charles:2004jd}
J.~Charles {\em et~al.} {\em Eur.Phys.J.} {\bf C41} (2005) 1--131,
  [\href{http://xxx.lanl.gov/abs/hep-ph/0406184}{{\tt hep-ph/0406184}}].
  updated results and plots available at: \url{http://ckmfitter.in2p3.fr}.

\bibitem{Bona:2006ah}
M.~Bona {\em et~al.} {\em JHEP} {\bf 10} (2006) 081,
  [\href{http://xxx.lanl.gov/abs/hep-ph/0606167}{{\tt hep-ph/0606167}}].
  {Standard Model fit results: Summer 2016 (ICHEP 2016):
  \url{http://www.utfit.org}}.

\bibitem{Tanabashi:2018oca}
M.~Tanabashi {\em et~al.} {\em Phys. Rev.} {\bf D98} (2018), no.~3 030001.
  \url{http://pdg.lbl.gov/2019/}.

\bibitem{Martinelli:2017}
G.~Martinelli {\em et~al.} \url{http://www.utfit.org/UTfit/}, 2017.

\bibitem{Buras2008:PhysRevD.78.033005}
A.~J. Buras and D.~Guadagnoli {\em Phys.Rev.} {\bf D78} (2008) 033005,
  [\href{http://xxx.lanl.gov/abs/0805.3887}{{\tt 0805.3887}}].

\bibitem{Brod2010:prd.82.094026}
J.~Brod and M.~Gorbahn {\em Phys.Rev.} {\bf D82} (2010) 094026,
  [\href{http://xxx.lanl.gov/abs/1007.0684}{{\tt 1007.0684}}].

\bibitem{Aoki:2019cca}
{\bf Flavour Lattice Averaging Group} Collaboration, S.~Aoki {\em et~al.}
  \href{http://xxx.lanl.gov/abs/1902.08191}{{\tt 1902.08191}}.

\bibitem{Jang:2015sla}
B.~J. Choi {\em et~al.} {\em Phys. Rev.} {\bf D93} (2016), no.~1 014511,
  [\href{http://xxx.lanl.gov/abs/1509.00592}{{\tt 1509.00592}}].

\bibitem{Blum:2014tka}
T.~Blum {\em et~al.} {\em Phys. Rev.} {\bf D93} (2016), no.~7 074505,
  [\href{http://xxx.lanl.gov/abs/1411.7017}{{\tt 1411.7017}}].

\bibitem{Laiho:2011np}
J.~Laiho and R.~S. Van~de Water {\em PoS} {\bf LATTICE2011} (2011) 293,
  [\href{http://xxx.lanl.gov/abs/1112.4861}{{\tt 1112.4861}}].

\bibitem{Durr:2011ap}
S.~Durr {\em et~al.} {\em Phys. Lett.} {\bf B705} (2011) 477--481,
  [\href{http://xxx.lanl.gov/abs/1106.3230}{{\tt 1106.3230}}].

\bibitem{Buras:2010}
A.~J. Buras, D.~Guadagnoli, and G.~Isidori {\em Phys.Lett.} {\bf B688} (2010)
  309--313, [\href{http://xxx.lanl.gov/abs/1002.3612}{{\tt 1002.3612}}].

\bibitem{Christ:2012}
N.~Christ {\em et~al.} {\em Phys.Rev.} {\bf D88} (2013), no.~1 014508,
  [\href{http://xxx.lanl.gov/abs/1212.5931}{{\tt 1212.5931}}].

\bibitem{Christ:2014qwa}
N.~Christ {\em et~al.} {\em PoS} {\bf LATTICE2013} (2014) 397,
  [\href{http://xxx.lanl.gov/abs/1402.2577}{{\tt 1402.2577}}].

\bibitem{Brod:2019rzc}
J.~Brod, M.~Gorbahn, and E.~Stamou
  \href{http://xxx.lanl.gov/abs/1911.06822}{{\tt 1911.06822}}.

\bibitem{Ben:2019}
B.~J. Choi {\em et~al.} {\em PoS} {\bf LATTICE2019} (2019) 050.

\bibitem{Seungyeob:2019}
S.~Jwa {\em et~al.} {\em PoS} {\bf LATTICE2019} (2019) 056.

\bibitem{Bhattacharya:2018ibo}
T.~Bhattacharya {\em et~al.} {\em PoS} {\bf LATTICE2018} (2018) 283,
  [\href{http://xxx.lanl.gov/abs/1812.07675}{{\tt 1812.07675}}].

\bibitem{Bailey:2017xjk}
J.~A. Bailey {\em et~al.} {\em EPJ Web Conf.} {\bf 175} (2018) 13012,
  [\href{http://xxx.lanl.gov/abs/1711.01786}{{\tt 1711.01786}}].

\bibitem{Bailey:2017zgt}
J.~Bailey, Y.-C. Jang, W.~Lee, and J.~Leem {\em EPJ Web Conf.} {\bf 175} (2018)
  14010, [\href{http://xxx.lanl.gov/abs/1711.01777}{{\tt 1711.01777}}].

\end{thebibliography}\endgroup


\end{document}